\documentclass[nofootinbib,showpacs]{revtex4}

\usepackage{bm}

\begin{document}

\title{Generating extremal neutrino mixing angles with Higgs family symmetries}

\author{Catherine I Low} \email{c.low@physics.unimelb.edu.au}
\affiliation{School of Physics, Research Centre for High Energy
  Physics, The University of Melbourne, Victoria 3010, Australia}
\date{\today}

\begin{abstract}
   The existence of maximal and minimal mixing angles in the neutrino
  mixing matrix motivates the search for extensions to the
  Standard Model that may explain these
  angles. A previous study (C.I.Low and R.R.Volkas,
  Phys.Rev.D68,033007(2003)), began a systematic search to find the minimal extension 
  to the Standard Model that explains these mixing angles. 
  It was found that in the minimal extensions to
  the Standard Model which allow neutrino oscillations, discrete
  unbroken lepton family symmetries only generate neutrino mixing
  matrices that are ruled out by experiment. This paper continues the 
  search by investigating all models with two or
  more Higgs doublets, and an Abelian family symmetry. It is found that discrete Abelian 
  family symmetries permit, but cannot explain, maximal 
  atmospheric mixing, however these models can ensure $\theta_{13}=0$.
\end{abstract}

\pacs{12.15.Ff, 14.60.Pq}

\maketitle

\section{Introduction}

The approximate form of the neutrino mixing matrix 
\begin{equation}\label{mnsform}
U_\mathrm{MNS}=\left(\begin{array}{ccc} 
\cos \theta_{12} & \sin \theta_{12} & 0 \\
-\frac{\sin \theta_{12}}{\sqrt{2}}&\frac{\cos \theta_{12}}{\sqrt{2}}&\frac{1}{\sqrt{2}}\\
\frac{\sin \theta_{12}}{\sqrt{2}}&-\frac{\cos \theta_{12}}{\sqrt{2}}&\frac{1}{\sqrt{2}}\end{array}\right),
\end{equation}
has been determined by neutrino oscillation experiments (Majorana
phases have not been included).

The mixing matrix, parameterised in the usual way, is formed by three
very different mixing angles.
The atmospheric mixing angle $\theta_{23}$ has a maximal value of $\pi/4$ at best fit 
\cite{SKatm,IMBatm,SOUDANatm,MACROatm}, the solar mixing angle $\theta_{12}$ has
been found to be large $\theta_{12} \approx 33^\circ$
\cite{fog03analysis}, but not maximal, by solar neutrino oscillation
experiments
\cite{HOMESTAKEsol,SAGEsol,GALLEXsol,SKsol,SNOsol,KAMLANDsol,GNOsol}, and the angle $\theta_{13}$,
measured by the non-observation of $\nu_e$ disappearance \cite{CHOOZ},
is small and has only an upper bound and is set to zero in Eq. (\ref{mnsform}). A special case of Eq. (\ref{mnsform}) is tri-bimaximal mixing, when $\sin \theta_{12}=\frac{1}{\sqrt{3}}$ \cite{hps1,zeehe1,xing,zeepar}, and $\theta_{13}$ is exactly zero. 
Two out of the three angles in Eq. (\ref{mnsform}) assume
extreme positions in parameter space -- the minimum possible value and
the maximum possible value -- so it has been suggested by many authors 
that this mixing pattern is not accidental, but could be due to a family
symmetry.

\subsection{Family symmetry models}

The symmetries of the Standard Model (SM) do not dictate the Yukawa coupling strength
between each fermion and the Higgs field. As a result, in the SM the charged
lepton and neutrino mass matrices are $3\times 3$ matrices with each
element a free variable. In the SM, Dirac mass matrices have nine free variables, and Majorana mass matrices have six. Diagonalising the mass matrices generates the
mixing matrix which can be any unitary $3\times 3$ matrix.  Family
symmetries constrain the form of the neutrino and charged lepton mass
matrices by relating elements of the mass matrix, or forcing elements
to be zero, thus reducing the number of free variables in the mass matrix.

For a family symmetry to fully predict a mixing matrix, all
mixing angles must be independent of the free variables in the mass
matrix, and must be prescribed by the form of the mass matrix. Mass
matrices that can generate mixing matrices in this approach have been
called ``form-diagonalisable'' matrices \cite{nogo1}.  
This can happen when there are three variables in the mass matrix corresponding to 
three unknown masses, and no free variables remaining for the mixing angles. An example of a form-diagonalisable matrix is the circulant matrix which can be generated by a $Z_3$ symmetry \cite{hps1}:
\begin{equation}\label{circ}
\left(\begin{array}{ccc}a&b&c\\c&a&b\\b&c&a\end{array}\right)
\end{equation}
is diagonalised by 
\begin{equation}
U=\frac{1}{\sqrt{3}}\left(\begin{array}{ccc}1&1&1\\1&\omega&\omega^*\\1&\omega^*&\omega\end{array}\right)
\end{equation}
where $\omega=e^{i 2\pi/3}$ and has eigenvalues $a+b+c$, $a+\omega b+\omega^* c$, $a+\omega^* b+\omega c$. 

The mixing matrix (Eq.  (\ref{mnsform})) may be created by partially
form-diagonalisable matrices, where the zero and maximal mixing angles
are not related to any free parameter, and arise from the form of the mass matrices, while the solar mixing angle may be related to a free parameter.

An Abelian family symmetry -- individual lepton number
$U(1)_{L_e}\times U(1)_{L_\mu}\times U(1)_{L_\tau}$ -- is conserved
when neutrinos are massless, but is broken when the three neutrinos gain 
different mass values. The special form of (\ref{mnsform}), and even more so the particular tri-bimaximal case makes it conceivable that a remnant of this Abelian group
remains unbroken with massive neutrinos, constraining the mixing
pattern. This further motivates the study of family symmetries, and Abelian symmetries in particular.

Many models with family symmetries have been proposed. A number of
these models \cite{hps1,hps2,maforminv} produce the desired form of
the mixing matrix, but use symmetries that cannot be easily
incorporated into the SM as the left-handed neutrinos transform in a
different way to the left-handed charged leptons, thus breaking
$SU(2)_L$. Most models that do preserve $SU(2)_L$ require additional
fields such as singlet or triplet Higgs fields \cite{glsinglets,
  maz2z3, zurab} or additional heavy fermions \cite{maa4}. The models with the least new particle content require a number
of Higgs doublets \cite{glz2,kubo,kubo2}, and some soft symmetry breaking terms to generate  Eq. (\ref{mnsform}).

It is clear that there are models that can produce Eq. (\ref{mnsform}),
the question that this work addresses is what is the minimal
predictive model. The approach taken is to construct the simplest model, and find out whether the model can generate the mixing matrix, or whether it can be ruled out. If it is ruled out, the next simplest model is investigated. 
A previous study \cite{nogo1} began a systematic search to find the
minimal extension to the SM that can generate the mixing
matrix form of Eq. (\ref{mnsform}), and found that for models with one SM Higgs doublet
unbroken discrete Abelian family symmetries cannot produce the matrix. In fact these symmetries can only generate mixing matrices that
are ruled out by experiment, or mixing matrices that are completely
unconstrained by the symmetry. Non-Abelian family symmetries are also ruled out as they dictate that the charged leptons are degenerate. 
The structure of the next simplest model is a subjective question. I chose to study extensions to the SM with two or more Higgs doublets that transform under an Abelian family symmetry. Abelian symmetries were chosen as the symmetry group could be a subgroup of $U(1)_{L_e} \times U(1)_{L_\mu} \times U(1)_{L_\tau}$, and for simplicity. This case also differs from the single Higgs doublet case as it is possible for the exact family symmetry to be spontaneously broken by the Higgs VEVs.

\subsection{Outline}

Section \ref{secconstraints} presents the mathematics
of family transformations, and shows how the mass and mixing matrices
can be constrained by the family symmetry transformations. Three neutrino mass generation mechanisms are considered; left and right handed neutrinos coupling to the Higgs doublets to create a Dirac neutrino mass matrix, left handed neutrinos coupling to the Higgs to form a dimension-5 operator, and the seesaw mechanism where the right handed neutrinos get a bare Majorana mass.
The types of mixing matrices that can be generated by an Abelian group
are described in section \ref{secmatrices}. I find that it is possible
for Abelian symmetries to dictate that $\theta_{13}=0$, and although the
symmetries permit all atmospheric mixing angles, the symmetries cannot specify that the atmospheric mixing angle is maximal.
Section \ref{secfavsym} lists group transformation matrices that give
a mixing matrix with $\theta_{13}=0$. Section \ref{secdiscuss} draws
conclusions about these models and suggests other models and
symmetries that may be more successful.

\section{How Higgs and lepton family symmetries constrain mass and mixing matrices}\label{secconstraints}

The following section describes how a single transformation can restrict the Higgs-lepton coupling matrices. For a symmetry group of order $n$ there are $n$ of these transformations. However, if the Lagrangian is unchanged by a transformation $X$, it will also be unchanged by $X^m$, where $m$ is a positive integer. For $Z_n$, the group of the addition of integers modulo $n$, the group is made up of the powers of one transformation, so a single transformation is sufficient to describe the restrictions placed on the coupling matrices. For all other groups more than one transformation is required.

\subsection{The symmetry transformations}

The family symmetry transformation matrices act on the different families of Higgs fields and leptons. The lepton
transformation is
\begin{equation}
\left(\begin{array}{c}\ell_L\\\nu_L\end{array}\right) \rightarrow X_L \left(\begin{array}{c}\ell_L\\\nu_L\end{array}\right), \qquad \ell_R\rightarrow X_{\ell R} \ell_R,\qquad \nu_R\rightarrow X_{\nu R} \nu_R,
\end{equation} 
where $\ell_L$, $\nu_L$, $\ell_R$ and $\nu_R$ are each 3-vectors in
family space, $\ell_L$ and $\ell_R$ are the vectors of left and right handed charged leptons, $\nu_L$ and $\nu_R$ are the vectors of left and right handed neutrinos. Each $X$ matrix is a $3\times 3$ unitary transformation matrix in
lepton family space. To preserve $SU(2)_L$, the left-handed neutrinos
transform in the same way as the left-handed charged leptons.

$n$ families of Higgs fields transform via
\begin{equation}
\Phi \rightarrow A_\Phi \Phi,
\end{equation}
where $\Phi=\left(\begin{array}{c}\phi_1 \\ \phi_2 \\ \vdots \\ \phi_n\end{array}\right)$ is an $n$-vector in Higgs family space, containing all the
Higgs fields, and $A_\Phi$ is an $n \times n$ unitary matrix.

\subsection{Constraints on the Higgs-lepton coupling matrices from the symmetry}

\subsubsection{The charged lepton Higgs coupling term}
The charged lepton-Higgs coupling transforms as
\begin{equation} 
 \overline{\ell_L} (\Phi^0)^T \lambda \ell_R \rightarrow \overline{\ell_L} X_L^\dagger (\Phi^0)^T  A_\Phi^T \lambda X_{\ell R} \ell_R,
\end{equation}
where $\Phi^0$ is an $n$-vector in Higgs family space containing just
the neutral component of the Higgs doublet. The term
$(\Phi^0)^T$ indicates a transpose in Higgs family space. $\lambda$ is
an $n$ vector in Higgs family space, where each element of
$\lambda$ is a $3\times 3$ Yukawa coupling matrix in lepton family
space.
Without any family symmetry the $\lambda$ matrices can be any $3\times
3$ matrices, but the existence of the family symmetry constrains them by
\begin{equation}\label{eqchrest}
\lambda = X_L^\dagger A_\Phi^T \lambda X_{\ell R}.
\end{equation}
Note that $X_L$ commutes with $\Phi$ as $X_L$ acts only on lepton
family space, and $\Phi$ is a lepton family singlet.  The
charged-lepton mass matrix is made up of the matrices in $\lambda$:
\begin{equation}
M_\ell =(<\Phi^0>)^T \lambda.
\end{equation}

\subsubsection{The Dirac neutrino Higgs coupling term}\label{dircoupling}

The Higgs field can couple to neutrinos in a number of ways.
A Dirac neutrino coupling term transforms like
\begin{equation}
\overline{\nu}_L(\Phi^0)^\dagger \kappa_{Dirac} \nu_R \rightarrow \overline{\nu}_L X_L^\dagger(\Phi^0)^\dagger A_\Phi^\dagger \kappa_{Dirac} X_{\nu R} \nu_R,
\end{equation}
where $\kappa_{Dirac}$ is an $n$-vector in Higgs family space, made up
of $3 \times 3$ Yukawa coupling matrices in lepton family space.
Without the symmetry, the Yukawa coupling matrices can be any $3
\times 3$ complex matrices.
Imposing the symmetry the matrices are constrained by
\begin{equation}\label{eqrestch}
\kappa_{Dirac} = X_L^\dagger A_\Phi^\dagger \kappa_{Dirac} X_{\nu R},
\end{equation}
which alters the Dirac neutrino mass matrix through
\begin{equation} 
M_{\nu Dirac}=(<\Phi^0>)^\dagger \kappa_{Dirac}. 
\end{equation}

\subsubsection{The dimension-5 neutrino Higgs coupling term}

A dimension-5 Higgs-neutrino coupling term transforms like
\begin{equation}
\frac{1}{\Lambda} \overline{\nu}_L\Phi^\dagger \kappa \Phi^* (\nu_L)^c \rightarrow
\frac{1}{\Lambda} \overline{\nu}_L X_L^\dagger \Phi^\dagger A_\Phi^\dagger \kappa A_\Phi^* \Phi^* X_L^* (\nu_L)^c,
\end{equation}
where $\kappa$ is now an $n\times n$ matrix in Higgs family space, with each component a $3\times 3$ symmetric matrix in lepton family space.
$\kappa$ is constrained by the symmetry through
\begin{equation}
\kappa =  X_L^\dagger A_\Phi^\dagger \kappa A_\Phi^* X_L^*,
\end{equation}
which consequently alters the mass matrix, defined by
\begin{equation}
M_\nu = (<\Phi^0>)^\dagger \kappa (<\Phi^0>)^*.
\end{equation}

\subsubsection{Higgs coupling terms for seesaw neutrinos}

In the seesaw mechanism the right-handed neutrinos couple to form a bare mass
term. The Higgs fields are not involved, so the mass term transforms as
\begin{equation}
\overline{\nu_R}^c  M_R \nu_R \rightarrow \overline{\nu_R}^c X_{\nu R}^T M_R X_{\nu R} \nu_R.
\end{equation}
restricting the heavy right-handed mass matrix $M_R$ by
\begin{equation}\label{eqrestmaj} 
M_R = X_{\nu R}^T M_R X_{\nu R}.
\end{equation}

The resultant light neutrino mass matrix, given by 
$M_\nu = M_{Dirac} M_R^{-1} M_{Dirac}^T$
is affected by the symmetry through the constraints on the heavy Majorana
mass matrix and the Dirac neutrino mass matrix (listed in Sub.Sec.
\ref{dircoupling}).

The seesaw case can be reduced to the dimension-5 operator case by relating
\begin{equation}
\kappa_{Dirac} M_R \kappa_{Dirac}^T = \kappa \textrm{ from dimension-5 case}.
\end{equation}
$\kappa_{Dirac} M_R \kappa_{Dirac}^T$ has all the constraints of
$\kappa$, plus additional restrictions from the transformation of
the right-handed neutrinos.

\subsection{Abelian groups create mass matrices with zero or unconstrained elements}\label{seczerounrest}

The restrictions family symmetries have on mass matrices depend on the transformation matrices that are chosen. If the set of matrices $X_i$ form a group then $Y_i=U^\dagger X_i U$ form the same group, where $U$ is any unitary matrix. Sets of matrices related in this way are called equivalent representations. Appendix \ref{appleptonsim} shows that choosing different equivalent representations for the lepton family symmetry transformations corresponds to choosing a different weak basis for the leptons. The constraints on the masses and on the mixing matrix are identical for two different equivalent representations. This makes it possible to eliminate groups, as each group has only a finite set of non-equivalent representations.  

This result simplifies the study of Abelian groups. As all Abelian groups are equivalent to a diagonal representation, only these representations need to be considered. Since the transformation matrices must be unitary, the diagonal elements are phases.

This makes the charged lepton restriction of Eq. (\ref{eqchrest}) become
\begin{equation}\label{eqchrestdiag}
\lambda_{ij}=(X_L^\dagger)_{ii} A_\Phi^T \lambda_{ij} (X_{\ell R})_{jj} \textrm{ no summation},
\end{equation}
where $i$, $j$ are lepton family indices. 

This restriction means that $\lambda^1_{ij}$ (the $ij$th element of
the Yukawa coupling matrix for $\phi_1$) can be related to the $ij$
element of the other Yukawa coupling matrices.  The symmetry however,
does not relate an element of $\lambda$ with a different element of
$\lambda$.  If the $ij$th elements of all $\lambda$ matrices are
zero then $M_{\ell ij}=0$, otherwise $M_{\ell ij}$ will most likely be
non-zero, and unrestricted by the symmetry. However, if two or more
$ij$th elements are related, and the Higgs VEVs are related
then there could be cancellation:
$M_{\ell ij}=\lambda_{ij} <\Phi^0>=0$. A relationship between VEVs is
possible as the symmetry also constrains the form of the Higgs
potential. If there isn't a
cancellation the element of the mass matrix is unrestricted by
the symmetry -- it is a free parameter of the model.

Consequently, a symmetry does not dictate the relationship between
any elements in the mass matrix. What the symmetry does do is force
some elements of the mass matrix to be zero, leaving all other
elements unrestricted. This is true for instances where there is
cancellation, and when there isn't a cancellation.  This makes the
analysis of Abelian groups easier, as only mass matrices with zero and
unrestricted elements need to be considered, and analysis of the Higgs
potential is not required.

There is only one Dirac mass matrix of this type that is
form-diagonalisable -- the diagonal matrix which is diagonalised by
the identity -- so most charged lepton diagonalisation matrices will
depend on the elements that are unrestricted by the symmetry.
Partially form-diagonalisable matrices are possible, for example a mass
matrix which is in $2\times 2$ block diagonal form, is diagonalised by
a unitary matrix which is in $2 \times 2$ block diagonal form, which
has one Euler angle depending on the free parameters, and the other
two angles zero.

Similarly, the Dirac neutrino-Higgs coupling matrices are restricted by a diagonal transformation,
\begin{equation}
\kappa_{Dirac \; ij} = (X_L^\dagger)_{ii} A_\Phi^\dagger \kappa_{Dirac \; ij} (X_{\nu R})_{jj},
\end{equation}
which also yields mass matrices with elements that are either zero or unrestricted.
The dimension-5 neutrino-Higgs coupling matrix is constrained by 
\begin{equation}\label{eqdim5diag}
\kappa_{ij} =  (X_L^\dagger)_{ii} A_\Phi^\dagger \kappa_{ij} A_\Phi^* (X_L^*)_{jj}.
\end{equation}
which also constrains some mass matrix elements to be zero. However, as Majorana mass matrices are symmetric, more form-diagonalisable mass matrices can be created. These mass matrices have a pseudo-Dirac form:
\begin{eqnarray}\label{pseudo1}
\left(\begin{array}{ccc}0&a&0\\a&0&0\\0&0&b\end{array}\right)
&\textrm{is diagonalised by}&
\left(\begin{array}{ccc}1/\sqrt{2}&-1/\sqrt{2}&0\\1/\sqrt{2}&1/\sqrt{2}&0\\0&0&1\end{array}\right)\equiv U_{\nu 1},
\\ \label{pseudo2}
\left(\begin{array}{ccc}a&0&0\\0&0&b\\0&b&0\end{array}\right)
&\textrm{is diagonalised by}&
\left(\begin{array}{ccc}1&0&0\\0&1/\sqrt{2}&-1/\sqrt{2}\\0&1/\sqrt{2}&1/\sqrt{2}\end{array}\right)\equiv U_{\nu 2},
\\ \label{pseudo3}
\left(\begin{array}{ccc}0&0&a\\0&b&0\\a&0&0\end{array}\right)
&\textrm{is diagonalised by}&
\left(\begin{array}{ccc}1/\sqrt{2}&0&-1/\sqrt{2}\\0&1&0\\1/\sqrt{2}&0&1/\sqrt{2}\end{array}\right)\equiv U_{\nu 3}.
\end{eqnarray}
The neutrinos that are mixed have $m_i=-m_j$.

Partially form-diagonalisable matrices also can be created:
\begin{eqnarray}\label{pseudo4}
\left(\begin{array}{ccc}0&a&b\\a&0&0\\b&0&0\end{array}\right)
&\textrm{is diagonalised by}&
\left(\begin{array}{ccc} -\frac{1}{\sqrt{2}}e^{-i \sigma} & \frac{1}{\sqrt{2}}e^{-i \sigma} & 0 \\ 
\frac{\sin \theta}{\sqrt{2}} e^{-i \sigma} & \frac{\sin \theta}{\sqrt{2}} e^{-i \sigma} & \cos \theta \\ 
\frac{\cos \theta}{\sqrt{2}} & \frac{\cos \theta}{\sqrt{2}} & -\sin \theta e^{i \sigma}
\end{array} \right)\equiv U_{\nu 4}, 
\\\label{pseudo5}
\left(\begin{array}{ccc}0&a&0\\a&0&b\\0&b&0\end{array}\right)
&\textrm{is diagonalised by}&
\left(\begin{array}{ccc} 
\frac{\sin \theta}{\sqrt{2}} e^{-i \sigma} & \frac{\sin \theta}{\sqrt{2}} e^{-i \sigma} & \cos \theta \\ 
 -\frac{1}{\sqrt{2}}e^{-i \sigma} & \frac{1}{\sqrt{2}}e^{-i \sigma} & 0 \\
\frac{\cos \theta}{\sqrt{2}} & \frac{\cos \theta}{\sqrt{2}} & -\sin \theta e^{i \sigma}
\end{array} \right)\equiv U_{\nu 5}, 
\\\label{pseudo6}
\left(\begin{array}{ccc}0&0&a\\0&0&b\\a&b&0\end{array}\right)
&\textrm{is diagonalised by}&
\left(\begin{array}{ccc} 
\frac{\cos \theta}{\sqrt{2}} & \frac{\cos \theta}{\sqrt{2}} & -\sin \theta e^{i \sigma} \\
\frac{\sin \theta}{\sqrt{2}} e^{-i \sigma} & \frac{\sin \theta}{\sqrt{2}} e^{-i \sigma} & \cos \theta \\ 
-\frac{1}{\sqrt{2}}e^{-i \sigma} & \frac{1}{\sqrt{2}}e^{-i \sigma} & 0 
\end{array} \right)\equiv U_{\nu 6}. 
\end{eqnarray}
where $\theta$ and $\sigma$ are angles which depend on the parameters $a$ and $b$.
Again, the two neutrinos that are maximally mixed have $m_i=m_j$.

Right handed Majorana mass matrices are constrained by a diagonal transformation by
\begin{equation}
M_{R\;ij} = (X_{\nu R}^T)_{ii} M_{R\;ij} X_{\nu R\;jj},
\end{equation}
and can also generate the form-diagonalisable matrices in equations (\ref{pseudo1} -- \ref{pseudo6}).

\subsection{Non-Abelian groups}

For family symmetries where the Higgs fields do not transform, non-Abelian symmetries ensure that at least two charged leptons must be degenerate \cite{nogo1}. In this case the mass matrix is constrained by the symmetry through the equation $M_\ell = X_L^\dagger M_\ell X_{\ell R}$. Consider the mass basis, where $M_\ell$ is diagonal. Non-Abelian representations cannot be equivalent to a diagonal representation, so the transformation associated with the mass basis will mix the mass matrix elements, and ensure that at least two of the masses are equal. For cases with more than one transforming Higgs field, the mass matrices are made up of a number of Yukawa matrices, and the transformations act on these matrices in a more complicated way than the one Higgs field case. Non-Abelian transformations no longer neccesarily force equal mass constraints, and as a result cannot be ruled out in the same way as in Ref \cite{nogo1}.

\section{Mixing angles that can be generated by Abelian groups}\label{secmatrices}

\subsection{Mass matrices to investigate}\label{secprogram}

To find the types of mixing matrices that can be created by Abelian groups a program was created to generate all sets of neutrino and charged lepton mass matrices with zero and unrestricted elements. 
For a given set of neutrino and charged lepton mass matrix types
two sets of mass matrices were created -- each with the same textures
(i.e. the same positions of the
zeros), but different random numbers were used for the elements that were
unrestricted. The unitary diagonalisation matrix was found for each
mass matrix, and two mixing matrices were created and compared. If an
angle was the same for both mixing matrices, the value of the angle 
was due to the textures of the mass matrices,
and thus, a result of the symmetry. If an angle was different for the two mixing matrices then the angle's value was due to the random numbers in the mass matrices, and not concerned with the symmetry. 

This was done for Dirac neutrinos, Majorana neutrinos which gained mass from a dimension-5 operator, and seesaw neutrinos. For seesaw neutrinos only right-handed Majorana neutrino mass matrices that were invertible were used, as non-invertible matrices generate less than three ultra-light neutrinos \cite{glasseesaw,fuku}.

\subsection{Results}\label{results}

\subsubsection{Abelian groups can generate zero and maximal mixing angles}

The only form-diagonalisable Dirac mass matrices that can be generated by Abelian groups create diagonalisation matrices with Euler angles equalling zero (Sec. \ref{seczerounrest}). Consequently, mixing angles for Dirac neutrinos are either zero, or unfixed by the symmetry, meaning Dirac neutrino models can ensure
$\theta_{13}=0$ but cannot fix the atmospheric or solar mixing
angles. These angles will be free parameters and can take any value. In some cases these two mixing angles can be related to lepton masses. 
 
Majorana neutrino mass matrices, from seesaw and dimension-5 operators, can
create fixed zero mixing angles, and also create fixed maximal mixing angles from the pseudo-Dirac type mass matrix (Eq. (\ref{pseudo1} -- \ref{pseudo6})). 
This looks promising for creating a maximal atmospheric mixing
angle, and the fact that maximal mixing angles can only be generated
from  Majorana matrices perhaps could be a key to explaining why
lepton mixing angles are large but quark mixing angles are
not. 
Unfortunately, it was found that the maximal mixing angle cannot correspond to
atmospheric mixing.

\subsubsection{Conditions for maximal mixing}

Mixing angles that are fixed to be maximal can only arise from the form-diagonalisable Majorana mass matrices listed in equations \ref{pseudo1} -- \ref{pseudo6}.
These  matrices have a $2 \times 2$ pseudo-Dirac block and the
maximally mixed neutrinos always have $m_i=-m_j$, corresponding to
$\delta m^2 =0$. This means that there will be no oscillation, however
a small mass squared difference could be created by  breaking the symmetry.

To demonstrate the difficulty in generating maximal atmospheric mixing, consider mixing matrices that have
$\theta_{13}=0$ and one maximal mixing angle. Matrices of this type
must have a pseudo-Dirac neutrino diagonalisation matrix (Eq. (\ref{pseudo1} -- \ref{pseudo6})) and a charged
lepton diagonalisation matrix that has either zero or one mixing
angle. That is,
\begin{equation}
U_{\ell 1}=I, \:\: U_{\ell 2}=\left(\begin{array}{ccc}
    \cos\alpha&-\sin\alpha&0\\ e^{i\delta} \sin \alpha & e^{i\delta} \cos \alpha&0 \\0&0&1
    \end{array}\right),
\:\:
U_{\ell 3}=\left(\begin{array}{ccc}1&0&0\\0&
    \cos\alpha&-\sin\alpha\\0&e^{i\delta}\sin\alpha & e^{i\delta} \cos \alpha\end{array}\right),
\: \:
U_{\ell 4}=\left(\begin{array}{ccc}\cos\alpha&0&-\sin\alpha\\0&1&0\\ e^{i\delta}\sin\alpha &0& e^{i\delta} \cos\alpha \end{array}\right),
\end{equation}
where $\alpha$ and $\delta$ are angles undefined by the symmetry.

Charged lepton diagonalisation matrices with more than one mixing angle cannot
produce mixing matrices with zero and maximal mixing
angles, 
so the only combinations of diagonalisation matrices that have both
$\theta_{13}=0$ and a maximal mixing angle are
\begin{eqnarray} 
U_{\mathrm{MNS} 1}&=&U_{\ell 1}^\dagger U_{\nu_1} =
\left(\begin{array}{ccc}1/\sqrt{2}&-1/\sqrt{2}&0\\1/\sqrt{2}&1/\sqrt{2}&0\\0&0&1\end{array}\right),
\\
U_{\mathrm{MNS} 2}&=&U_{\ell 1}^\dagger U_{\nu_2} = \left(\begin{array}{ccc}1&
    0&0\\0&1/\sqrt{2}&-1/\sqrt{2}\\0&1/\sqrt{2}&1/\sqrt{2}\end{array}\right),
\\
U_{\mathrm{MNS} 3}&=&U_{\ell 1}^\dagger U_{\nu 4}=U_{\ell 3}^\dagger U_{\nu 4}=U_{\ell 3}^\dagger U_{\nu 1}=\left(\begin{array}{ccc} -\frac{1}{\sqrt{2}} & \frac{1}{\sqrt{2}} & 0 \\ 
\frac{\sin \theta}{\sqrt{2}} & \frac{\sin \theta}{\sqrt{2}} & \cos \theta \\ 
\frac{\cos \theta}{\sqrt{2}} & \frac{\cos \theta}{\sqrt{2}} & -\sin \theta \\
\end{array} \right).
\end{eqnarray}
where the $U_{\nu i}$'s are given in Eqs: (\ref{pseudo1}-\ref{pseudo6})

The only mixing matrix that has maximal atmospheric mixing is
$U_{\mathrm{MNS} 2}$, which also has a very unsatisfactory solar
mixing angle of zero. $U_{\mathrm{MNS} 3}$ has maximal solar mixing, but the
atmospheric angle is not dictated by the symmetry.

The program that was written (see Sub.Sec. \ref{secprogram}) also searched for fixed mixing angles without the  
$\theta_{13}=0$ constraint and found that $\theta_{23}$ is still unfixed by the symmetry, therefore the only aspect of the mixing matrix form of Eq.
(\ref{mnsform}) that can be generated by a symmetry is $\theta_{13}=0$. It is not possible to demonstrate this result in a concise way in this paper: it is instead the result of a systematic computer-aided search.
We have seen that the mixing angles can be zero, maximal or unfixed by the symmetry. The unfixed mixing angles can be either related to fermion masses or completely free variables. The possible ways in which mixing angles can be related to masses was not analysed by the program, however, for the cases where $\theta_{13}=0$ mass-mixing angle relationships were worked out by hand (Sub.Sec.\ref{secmassmixing}) .  

Note that the fact that solar mixing can be forced to be maximal and atmospheric
mixing angle cannot be forced to be maximal does not indicate any fundamental
difference between the flavours. If one mixing matrix can be predicted
by an Abelian group, so can the mixing matrix with rows permuted.
Permuting the rows corresponds to interchanging $e$, $\mu$ or $\tau$, so
the whole set of possible neutrino mixing matrices are flavour symmetric.

The mixing angles, defined as Euler angles, however are not flavour symmetric. The probability that a neutrino of flavour $\ell$ is detected as flavour $\ell'$ after a distance $x$ is given by
\begin{equation}
P_{\ell \rightarrow \ell'}(x)=\sum_m U_{\ell m}^2 U_{\ell' m}^2+\sum_{m'\neq m} U_{\ell m} U_{\ell m'} U_{\ell' m'} U_{\ell' m} \cos \left(2 \pi \frac{x}{L_{m m'}}\right)
\end{equation}
where $L_{m m'}=2 \pi \frac{2 p_nu}{\delta m_{m m'}^2}$.
 When $\theta_{13}=0$, the probability of an electron neutrino being detected as an electron neutrino after a distance $x$ is
\begin{equation}
P_{e \rightarrow e}(x)=1-\sin ^2 2 \theta_{12} \sin^2 \left(\frac{\pi x}{L_{12}}\right),
\end{equation}
and is only dependent on $\theta_{12}$. The probability of a muon neutrino being detected as a muon neutrino after a distance $x$,
\begin{eqnarray}
P_{\mu \rightarrow \mu}(x)&=&\sin^4 \theta_{12} \cos^4 \theta_{23} + \cos^4 \theta_{12}\cos \theta^4_{23} + \sin^4 \theta_{23} 
+ 2 \sin^2 \theta_{12} \cos^2 \theta_{12} \cos^4 \theta_{23} \cos\left(2 \pi \frac{x}{L_{12}}\right) \\
& & +  2 \sin^2 \theta_{12} \cos^2 \theta_{23}\sin^2 \theta_{23} \cos\left(2 \pi \frac{x}{L_{13}}\right)
+  2 \cos^2 \theta_{12}\cos^2 \theta_{23} \sin^2 \theta_{23} \cos\left(2 \pi \frac{x}{L_{23}}\right),
\end{eqnarray}
is dependent on both non-zero mixing angles. So when $\theta_{13}=0$ a maximal solar mixing angle corresponds to a maximum amplitude of oscillation -- an electron neutrino will oscillate into a state with no electron neutrino component. 
Maximal atmospheric mixing means that
there is a mass eigenstate that is an equal superposition of $\nu_\mu$
and $\nu_\tau$ and does not imply a maximum amplitude of oscillation in the three flavour case.

\section{Mass matrices and symmetries that produce $\theta_{13}=0$}\label{secfavsym}

There are several sets of charged lepton and neutrino mass matrices
that can produce $\theta_{13}=0$, some of which can be created from a
symmetry. The ones that can be related to a symmetry, and do not force the muon or tau leptons to be massless are listed in tables \ref{tabmaj} and \ref{tabdirac}, along with the smallest symmetry group that can produce the mass matrices. All cases that can be generated by a symmetry require two Higgs doublets, unless otherwise stated in the table. Cancellation within the mass matrix was not considered, nor was the possibility of VEVs equalling zero (i.e. if $M^{ij}_\ell=0$, then it was assumed that $\lambda^{ij}_{1,2,...,n}=0$).
With these assumptions, diagonal Higgs transformations give the same mixing matrices as equivalent non-diagonal representations (see App. \ref{apphiggssim}), so to find the smallest symmetry group only diagonal transformations were investigated. It is possible that a smaller group than that listed could produce the mixing matrices if there are cancellations or zero VEVs. 

The smallest group that can give $\theta_{13}=0$ is $Z_3$, the group of addition modulo 3. 
$Z_2$ gives the same
mixing matrices that can be generated in the single Higgs doublet
case, as analysed in Ref. \cite{nogo1} and are either unrestricted or experimentally ruled out.
This is shown in App. \ref{z2ruledout}.

\begin{table}
  \begin{center}
    \begin{tabular}{|l|l|l|l|l|}
      \hline
& $M_\nu$ & $M_\ell$ & Smallest Symmetry &  Mass Restrictions \\
      \hline \hline       
1&$\left(\begin{array}{ccc} A & B & 0 \\ B & D & 0 \\0 & 0 & F \end{array}
\right)$
&
$\left(\begin{array}{ccc} a & 0 & 0 \\ 0 & d & e \\0 & f & g \end{array}
\right)$
&
\begin{tabular}{l}Dimension-5: 3 Higgs doublets, $Z_7$ \\ Seesaw: $Z_4$ \end{tabular}
& 
No mass restrictions
\\
\hline
2&$\left(\begin{array}{ccc} A & B & 0 \\ B & 0 & 0 \\0 & 0 & F \end{array}
\right)$
&
$\left(\begin{array}{ccc} a & 0 & 0 \\ 0 & d & e \\0 & f & g \end{array}
\right)$
&
\begin{tabular}{l}Dimension-5: $Z_5$ \\ Seesaw: $Z_4$ \end{tabular}
& 
\begin{tabular}{l} $\theta_{12}$ is related to neutrino masses \\ giving neutrino mass hierarchy. \\ \end{tabular} 
\\
\hline
3&$\left(\begin{array}{ccc} 0 & A & 0 \\ A & B & 0 \\0 & 0 & C \end{array}
\right)$
&
$\left(\begin{array}{ccc} a & 0 & 0 \\ 0 & d & e \\0 & f & g \end{array}
\right)$
&
\begin{tabular}{l}Dimension-5: 3 Higgs doublets $Z_9$ \\ Cannot be generated with seesaw neutrinos \end{tabular}
& 
\begin{tabular}{l} $\theta_{12}$ is related to neutrino masses \\ giving neutrino mass hierarchy. \\ \end{tabular} 
\\
\hline
4&$\left(\begin{array}{ccc} A & B & C \\ B & 0 & 0 \\C & 0 & 0 \end{array}
\right)$  
& 
$ \left(\begin{array}{ccc} a & 0 & 0 \\ 0 & d & e \\0 & f & g \end{array}
\right)$   
& 
\begin{tabular}{l}Dimension-5: $Z_5$\\ Seesaw: $Z_5$ \end{tabular}
& 
\begin{tabular}{l}$\theta_{12}$ related to neutrino masses \\ giving nearly maximal solar mixing \\ therefore ruled out by experiment \\ \end{tabular}    \\
\hline
5&$\left(\begin{array}{ccc} A & B & 0 \\ B & D & 0 \\0 & 0 & F \end{array}
\right)$
&
$\left(\begin{array}{ccc} 0 & 0 & 0 \\ a & b & c \\d & e & f \end{array}
\right)$
&
\begin{tabular}{l}Dimension-5: 3 Higgs doublets required $Z_7$\\ Seesaw: $Z_4$ or $Z_3$ if $m_3=0$ \end{tabular}
&  
\begin{tabular}{l}$m_e=0$\\  \end{tabular}\\
\hline
6&$\left(\begin{array}{ccc} A & B & 0 \\ B & 0 & 0 \\0 & 0 & C \end{array}
\right)$
&
$\left(\begin{array}{ccc} 0 & 0 & 0 \\ a & b & c \\d & e & f \end{array}
\right)$
&
\begin{tabular}{l}Dimension-5: $Z_5$\\ Seesaw: $Z_5$ \end{tabular}
&
\begin{tabular}{l}$m_e=0$ \\$\theta_{12}$ related to neutrino masses \\ giving a hierarchical neutrino mass pattern \end{tabular}
\\
\hline
7&
$\left(\begin{array}{ccc} 0 & A & 0 \\ A & B & 0 \\0 & 0 & C \end{array}
\right)$
&
$\left(\begin{array}{ccc} 0 & 0 & 0 \\ a & b & c \\d & e & f \end{array}
\right)$
&
\begin{tabular}{l}Dimension-5: 3 Higgs doublets $Z_9$\\ Cannot be generated with seesaw neutrinos \end{tabular}
&
\begin{tabular}{l}$m_e=0$ \\$\theta_{12}$ related to neutrino masses \\ giving a hierarchical neutrino mass pattern \end{tabular}
\\
\hline
8&$\left(\begin{array}{ccc} A & B & C \\ B & 0 & 0 \\C & 0 & 0 \end{array}
\right)$  
& 
$ \left(\begin{array}{ccc} 0 & 0 & 0 \\ a & b & c \\d & e & f \end{array}
\right)$   
&
\begin{tabular}{l}Dimension-5: $Z_5$\\ Seesaw: $Z_5$ \end{tabular}
&
\begin{tabular}{l}$\theta_{12}$ related to neutrino masses \\ giving nearly maximal solar mixing \\ therefore ruled out by experiment \\ \end{tabular}    \\
\hline
9&$\left(\begin{array}{ccc} A & B & 0 \\ B & D & 0 \\0 & 0 & F \end{array}
\right)$
&
$\left(\begin{array}{ccc} 0 & 0 & a \\ 0 & 0 & b \\c & d & e \end{array}
\right)$
&
\begin{tabular}{l}Dimension-5: $Z_4$\\Seesaw: $Z_4$ \end{tabular}
&
\begin{tabular}{l}$m_e=0$ \\ \end{tabular}    
\\
\hline
10 &$\left(\begin{array}{ccc} A & B & 0 \\ B & D & 0 \\0 & 0 & F \end{array}
\right)$ 
& 
$\left(\begin{array}{ccc} 0 & a & 0 \\ 0 & b & 0 \\0 & c & d \end{array}\right) $ 
& 
\begin{tabular}{l}Dimension-5: $Z_4$\\ Seesaw: $Z_4$ \end{tabular}
& 
 \begin{tabular}{l}$m_e=0$  \end{tabular}
\\ \hline 
    \end{tabular}
  \end{center}
  \caption{Mass matrices for Majorana neutrinos that give $\theta_{13}=0$. Two Higgs doublets are required unless otherwise stated. }
  \label{tabmaj}
\end{table}

\begin{table}
  \begin{center}
    \begin{tabular}{|l|l|l|l|l|}
      \hline
    &  $M_\nu$ & $M_\ell$ & Smallest Symmetry &  Mass Restrictions \\
      \hline \hline       
1&$\left(\begin{array}{ccc} A & B & 0 \\ C & D & 0 \\0 & 0 & E \end{array}
\right)$
&
$\left(\begin{array}{ccc} a & 0 & 0 \\ 0 & b & c \\0 & d & e \end{array}
\right)$
&
\begin{tabular}{l}  $Z_4$ \end{tabular}
& 
No mass restrictions
 \\
\hline
2&$\left(\begin{array}{ccc} A & B & C \\ D & E & F \\0 & 0 & 0 \end{array}
\right)$
&
$\left(\begin{array}{ccc} a & 0 & 0 \\ 0 & b & c \\0 & d & e \end{array}
\right)$
&
\begin{tabular}{l}  $Z_4$ \end{tabular}
& 
$m_3=0$
 \\
\hline
3&$\left(\begin{array}{ccc} A & B & 0 \\ C & D & 0 \\0 & 0 & E \end{array}
\right)$
&
$\left(\begin{array}{ccc} 0 & 0 & 0 \\ a & b & c \\d & e & f \end{array}
\right)$
&
\begin{tabular}{l}  $Z_4$ \end{tabular}
&  
$m_e=0$\\
\hline
4&$\left(\begin{array}{ccc} A & B & C \\ D & E & F \\0 & 0 & 0 \end{array}
\right)$
&
$\left(\begin{array}{ccc} 0 & 0 & 0 \\ a & b & c \\d & e & f \end{array}
\right)$
&
\begin{tabular}{l}  $Z_3$ \end{tabular}
&  
\begin{tabular}{l}$m_e=0$ and $m_3=0$ \\  \end{tabular}\\
\hline
5&$\left(\begin{array}{ccc} A & B & 0 \\ C & D & 0 \\0 & 0 & E \end{array}\right)$ & $\left(\begin{array}{ccc} 0 & 0 & a \\ 0 & 0 & b \\c & d & e \end{array}
\right)$ 
&  
\begin{tabular}{l}  $Z_4$ \end{tabular}
&
$m_e=0$
\\
\hline
6&$\left(\begin{array}{ccc} A & B & C \\ D & E & F \\0 & 0 & 0 \end{array}\right)$ 
& $\left(\begin{array}{ccc} 0 & 0 & a \\ 0 & 0 & b \\c & d & e \end{array}
\right)$ 
&  
\begin{tabular}{l}  $Z_4$ \end{tabular}
& \begin{tabular}{l}$m_e=0$ and $m_3=0$ \\  \end{tabular}\\
\hline
7& $\left(\begin{array}{ccc} A & B & 0 \\ C & D & 0 \\0 & 0 & E \end{array}
\right)$ 
& 
$\left(\begin{array}{ccc} 0 & a & 0 \\ 0 & b & 0 \\0 & c & d \end{array}\right) $ 
& 
\begin{tabular}{l}  $Z_4$ \end{tabular}
&
\begin{tabular}{l}$m_e=0$ \\  \end{tabular}\\
 \hline
8& $\left(\begin{array}{ccc} A & B & C \\ D & E & F \\0 & 0 & 0 \end{array}
\right)$ 
& 
$\left(\begin{array}{ccc} 0 & a & 0 \\ 0 & b & 0 \\0 & c & d \end{array}\right) $ 
& 
\begin{tabular}{l}  $Z_4$ \end{tabular}
 & \begin{tabular}{l}$m_e=0$ and $m_3=0$ \\  \end{tabular}
\\ \hline
    \end{tabular}
  \end{center}
  \caption{Mass matrices for Dirac neutrinos that give $\theta_{13}=0$. Transpositions of the columns of the mass matrices do not alter the masses, the mixing matrix, or the symmetry. Two Higgs doublets are required unless otherwise stated.}
  \label{tabdirac}
\end{table}

\subsection{Examples of symmetry transformations}

For seesaw neutrinos the mass matrices in the first row of table \ref{tabmaj} can be generated by the $Z_4$ transformation
\begin{equation}
A_\Phi=\left(\begin{array}{cc}i & 0 \\ 0 & -1\end{array}\right),\qquad 
X_L=\left(\begin{array}{ccc}i & 0 &0\\ 0 & 1 &0 \\ 0&0&-i \end{array}\right),
\qquad
X_{\ell R}=X_{\nu R}=\left(\begin{array}{ccc}1 & 0 &0\\ 0 & -1 &0 \\ 0&0&-1 \end{array}\right).
\end{equation}

The first row of mass matrices in table \ref{tabdirac} can be generated by the $Z_4$ transformation
\begin{equation}
A_\Phi=\left(\begin{array}{cc}1 & 0 \\ 0 & i\end{array}\right), \qquad 
X_L=\left(\begin{array}{ccc}i & 0 &0\\ 0 & 1 &0 \\ 0&0&-i \end{array}\right),
\qquad
X_{\ell R}=\left(\begin{array}{ccc}i & 0 &0\\ 0 & -i &0 \\ 0&0&-i \end{array}\right),
\qquad
X_{\nu R}=\left(\begin{array}{ccc}i & 0 &0\\ 0 & i &0 \\ 0&0&-i \end{array}\right).
\end{equation}

For many cases the Dimension-5 mass matrices require larger symmetries or additional Higgs doublets. This is because the seesaw and Dirac cases have extra freedom due to the transformation of the right-handed neutrino.

The first row of matrices in table \ref{tabmaj} can be generated by a $Z_7$ transformation involving three Higgs doublets:
\begin{equation}
A_\phi=\left(\begin{array}{ccc}1 & 0 &0 \\ 0&\beta^2 &0 \\0&0&\beta \end{array}\right) \qquad X_L=\left(\begin{array}{ccc}\beta^6&0&0\\0&1&0\\0&0&\beta^2 \end{array}\right) \qquad
X_{\ell R}=\left(\begin{array}{ccc}\beta^6&0&0\\ 0&1&0\\0&0&1\end{array}\right)
\end{equation}
where $\beta=e^{2 \pi i/7}$.

\subsection{Mass and mixing angle relationships}\label{secmassmixing}

Section \ref{results} established that none of the mass matrices that give $\theta_{13}=0$ can give a fixed atmospheric mixing angle. In fact, it can be shown by diagonalising the mass matrices of tables \ref{tabmaj} and \ref{tabdirac}, that the atmospheric mixing angle is also unrelated to masses -- it is a free variable for all $\theta_{13}=0$ cases. 

In cases where a Majorana mass matrix is the source for the solar mixing angle, the mixing angle can relate to the masses.
These cases are the Majorana neutrino matrices from rows 2 -- 4 and 5 -- 8 of table \ref{tabmaj}.

The neutrino mass matrix of rows 2 and 6 of table \ref{tabmaj} 
\begin{equation}
\left(\begin{array}{ccc}A & B &0\\ B & 0 &0 \\ 0&0&D \end{array}\right),
\end{equation}
 relate the masses and mixing angle by
\begin{equation}
\tan 2\theta_{sol}=\frac{2\sqrt{-m_1 m_2}}{m_1+m_2}.
\end{equation}
 Using the approximate values for $\delta m^2_{12} \approx 7.5\times 10^{-5} eV^2$ and $\theta_{sol}\approx 33^\circ$, the neutrino masses must be approximately hierarchical, with $|m_1|\sim 9 \times 10^{-3} eV$, $|m_2|\sim 4 \times 10^{-3} eV$, and $|m_3|\sim 0.04 \; eV$.  
Rows 4 and 7 give the same mass pattern except $m_1$ and $m_2$ are interchanged.  

The mass matrices of rows 5 and 8 are even more constrained. There are only 3 free variables in these neutrino mass matrices, and these free variables describe two masses, the solar mixing angle, and a contribution to the atmospheric mixing angle. As a result the mixing angles and masses must be related: 
\begin{eqnarray}
m_1&=&\frac{1}{2}(|A|\pm\sqrt{|A|^2+4(|B|^2+|C|^2)}),\\
m_2&=&\frac{1}{2}(|A|\mp\sqrt{|A|^2+4(|B|^2+|C|^2)}),\\
m_3&=&0,\\
\tan(2 \theta_{12})&=&\frac{2 \sqrt{|B|^2+|C|^2}}{A},\\
\tan \theta_{23\nu}&=&\frac{|C|}{|B|}, 
\end{eqnarray}
where $\theta_{23\nu}$, along with the diagonalisation angle from the charged lepton mixing matrix form the atmospheric mixing angle. 
 To achieve $\delta m^2_{12} \ll \delta m^2_{23}$, $|A| \ll \sqrt{|B|^2+|C|^2}$ is required, and the neutrino mass matrix becomes close to the partially form-diagonalisable matrix of Eq. \ref{pseudo4}, giving very nearly maximal solar mixing. This has been ruled out by experiment.

\subsection{Flavour changing neutral currents}

Models involving a number of Higgs doublets and
Abelian symmetries naturally predict flavour changing neutral currents
(FCNCs) for charged leptons and neutrinos \cite{gwfcnc}, and the models presented in this paper are likely to be no exception. However it is possible that the flavour symmetry somehow suppresses the FCNCs. Many of the charged lepton mass matrices listed in tables \ref{tabmaj}  and \ref{tabdirac} only mix  $\mu$ and $\tau$, and do not mix electrons, meaning that the most experimentally constrained FCNC processes (such as $\mu \rightarrow \overline{e} ee $) are not allowed. However the decay $\tau \rightarrow \overline{\mu} \mu \mu$
is allowed, and as large off diagonal elements in the Yukawa
coupling matrices are required to give large mixing angles, this transition is not likely suppressed. However, it is possible that some action of the flavour symmetry on the Higgs fields can prevent the FCNCs from becoming too large.

\section{Conclusion}\label{secdiscuss}

The best fit neutrino mixing matrix Eq. (\ref{mnsform}) has a few
peculiar aspects; it is very different from the CKM matrix, and it has
one maximal mixing angle and one minimal mixing angle. It would be
pleasing to find that this pattern can be generated by a symmetry.
  
Earlier work \cite{nogo1} showed that unbroken lepton family symmetries
alone can only produce mixing matrices which are not allowed experimentally. This paper continues the search for a symmetry explanation
to the form of the mixing matrix. The models considered are extensions
of the SM that include a number of Higgs doublets and
discrete Abelian symmetries that transform the Higgs and lepton families.

Symmetries of this type can only fix mixing angles to be zero or
maximal, otherwise the angle can be any value as it depends on the free parameters of the model.
This is due to the fact that all Abelian representations are
equivalent to diagonal representations. In the diagonal basis Abelian
symmetries can only dictate whether an element in a mass matrix is
zero or unrestricted; no relationships between mass matrix elements
can be generated, so only a few form-diagonalisable mass matrices can
be generated, and most mixing angles are not fixed by the
symmetry. A small number of Majorana mass matrices can generate
maximal mixing, however, this mixing cannot correspond to the
atmospheric mixing angle. The characteristic of Eq. (\ref{mnsform}) that can be produced by Abelian family symmetries is $\theta_{13}=0$. This requires at least two Higgs doublets and a $Z_3$ or larger family symmetry.

Although Abelian symmetries have limited ability in predicting fixed mixing angles, symmetries can relate lepton masses and mixing angles. For the cases where $\theta_{13}$ is forced to be zero, the solar mixing angle can be related to neutrino masses. This relationship fixes the neutrino mass pattern to be hierarchical.

Non-Abelian family symmetries may produce better results,
as they can relate different elements in the mass matrices together,
possibly creating form-diagonalisable matrices that cannot be generated 
with Abelian groups. Extending the Higgs sector by including triplet Higgs fields to
generate neutrino mass is also likely to increase the possible types of mixing matrices.

The approach taken here, to find the minimal model that explains the neutrino 
mixing matrix, has succeeded in explaining one of the interesting aspects of the
mixing. However, the models that can explain $\theta_{13}=0$ are not particularly simple. Fixing this one variable requires the introduction of extra Higgs doublets, which additionally can lead to flavour changing neutral currents.
These results suggest that the minimal model route may not readily yield a satisfactory explanation for the mixing parameters. This could be a consequence of considering neutrino mixing independently of other unresolved issues in particle physics, such as mass hierarchy and quark mixing. Perhaps the answer can only be found by finding a model that simultaneously addresses several of these problems.

\acknowledgments{The author would like to thank R. Volkas for the initial inspiration for this project and many invaluable discussions, and C. Looser for programming assistance. This work was supported by The University of Melbourne.}

\appendix 

\section{Equivalent
  representations of the lepton transformations yield identical predictions}\label{appleptonsim}

Two equivalent representations for the fermion transformations, $A_L,A_{\ell R},A_{\nu R}$ and $B_L,B_{\ell R},B_{\nu R}$, are related by 
\begin{eqnarray}
B_L&=&S_L^\dagger A_L S_L, \\
B_{\ell R}&=&S_{\ell R}^\dagger A_{\ell R} S_{\ell R},\\
B_{\nu R}&=&S_{\nu R}^\dagger A_{\nu R} S_{\nu R}.
\end{eqnarray}
where $S_L,S_{\ell R}$ and $S_{\nu R}$ can be any $3\times 3$ unitary matrices. 

\subsection{Charged leptons}

The Yukawa matrices arising from the $B$ transformations are
denoted by a $B$ subscript. 
The restrictions from the $B$ transformations on the charged lepton Yukawas (from Eq. (\ref{eqrestch})) are
\begin{eqnarray}
\lambda_B&=&B_L^\dagger A_\Phi^T \lambda_B B_{\ell R} \\
&=& S_L^\dagger A_L^\dagger S_L A_\Phi^T \lambda_B S_{\ell R}^\dagger A_{\ell R} S_{\ell R}.
\end{eqnarray}
Rearranging gives
\begin{eqnarray}
S_L \lambda_B S_{\ell R}^\dagger &=& A_L^\dagger S_L A_\Phi^T \lambda_B S_{\ell R}^\dagger A_{\ell R} \\
&=& A_\Phi^T A_L^\dagger (S_L \lambda_B S_{\ell R}^\dagger) A_{\ell R}.
\end{eqnarray}
($X_\Phi^T$ commutes with $U_L^\dagger S_L$, as they operate in different spaces).

The charged lepton Yukawa restrictions for $A$ transformations are
$\lambda_A=A_\Phi^T A_L^\dagger \lambda_A A_{\ell R}$.
$S_L \lambda_B S_{\ell R}^\dagger$ has the same restrictions from the symmetry as $\lambda_A$. As the mass matrices are completely unconstrained
apart from the generation symmetry constraints, we can set $S_L \lambda_B S_{\ell R}^\dagger=\lambda_A$.
This means that the charged lepton mass matrix from the $A$ representation can be given by
\begin{equation}
M_{\ell A} = \lambda_A <\Phi>^T=
S_L \lambda^B S_{\ell R}^\dagger<\Phi>^T=
S_L M_{\ell B} S_{\ell R}^\dagger.
\end{equation}
If $M_{\ell A}$ is diagonalised by $U^A_{\ell L}$, and $U^A_{\ell R}$, then $M_{\ell B}$ will be diagonalised by $U^B_{\ell L}=S_L U^A_{\ell L}$ and $U^A_{\ell R}=S_{\ell R} U^A_{\ell R}$.

\subsection{Dimension-5 neutrino masses}\label{appdim5}

Similarly, the conditions on the dimension-5 neutrino coupling matrices mean we can identify
\begin{equation}
\kappa_A=S_L \kappa_B S_L^T,
\end{equation}
and the two mass matrices can be related by
\begin{equation}
M_{\nu A} = <\Phi>^\dagger \kappa_A <\Phi_i>^* =
<\Phi>^\dagger S_L \kappa_B S_L^T <\Phi_i>^* =S_L <\Phi>^\dagger \kappa_B <\Phi_i>^*S_L^T= S_L M_{\nu B} S_L^T. 
\end{equation}

$M_{\nu A}$ is diagonalised by  $U_{\nu A}$, and $M_{\nu B}$ is diagonalised by  $U_{\nu B}=S_L U_{\nu B}$,
 giving identical mixing matrices for the two transformations,
\begin{equation}
U_{\mathrm{MNS} B}=U^{B\dagger}_{\ell_L}U^B_{\nu}
=U_{\ell_L}^{A \dagger} S_L^\dagger S_L U_{\nu}=U_{\mathrm{MNS}A}.
\end{equation}

\subsection{Dirac neutrinos}

The neutrino Yukawa restrictions for the two transformations are related by 
$S_L \kappa_{Dirac B} S_{\nu R}^\dagger=\kappa_{Dirac A}$, and the two mass matrices can be related by $S_L M_{\nu B} S_{\nu R}^\dagger=M_{\nu_A}$. This gives diagonalisation matrices related by $U^{B}_{\nu L}=S_L U^A_{\nu L}$.
The mixing matrix is, therefore, the same for both transformations:
\begin{equation}
U^B_\mathrm{MNS}=U^{B\dagger}_{\ell_L}U^{B}_{\nu L}=U_{\ell_L}^{A \dagger} S_L^\dagger  S_L U^A_{\nu L}=U^A_\mathrm{MNS}.
\end{equation}

\subsection{Seesaw neutrinos}

The right-handed Majorana mass matrix is restricted by Eq. (\ref{eqrestmaj}), giving
\begin{equation}
M_{R B}=B^T_{\nu R} M_{R B} B_{\nu R}=
S_{\nu R}^T A_{\nu R}^T S_{\nu R}^* M_{R B} S_{\nu R}^\dagger A_{\nu R} S_{\nu R}, 
\end{equation}
while $M_{R A}$ is constrained by $M_{R A}=A^T_{\nu R} M_{R A} A_{\nu R}$. $M_{R A}$ can be equated to $S_{\nu R} M_{R B} S_{\nu R}^T$, as they have the same constraints from the symmetry.
 
The Dirac neutrino mass is as above: $S_L M_{Dirac B} S_{\nu R}^\dagger=M_{Dirac A}$. The resultant neutrino mass matrix is 
\begin{equation}
M_{\nu A}=M_{Dirac A}M_R^{-1} M_{Dirac A}^T=S_L M_{Dirac B}  M_{R B} M_{Dirac B}^* S_L^T=S_L M_{\nu B} S_L^T.
\end{equation}
The seesaw mass matrices are related to each other in the same way as the dimension-5 mass matrices. Using the result from Sub.Sec.\ref{appdim5}, the mixing matrices from the two representations are equal.

\section{Equivalent representations for the Higgs transformation give the
  same mixing predictions in most cases}\label{apphiggssim}

Two equivalent representations for the Higgs transformations $A_\Phi$ and $B_\Phi$, are related by 
\begin{equation}
B_\Phi=S^\dagger A_\Phi S.
\end{equation}
The Yukawa matrices arising from the $A$ and $B$ transformation are
denoted by an $A$ or $B$ subscript. 

The charged lepton Yukawa restrictions for the $B$ transformation are
\begin{eqnarray}
\lambda_B&=&X_L^\dagger B_\Phi^T \lambda_B X_{\ell R} \\
&=& X_L^\dagger S^T A_\Phi^T S^*\lambda_B  X_{\ell R}. 
\end{eqnarray}
Rearranging gives
\begin{eqnarray}
S^* \lambda_B &=& X_L^\dagger A_\Phi^T S^*\lambda_B X_{\ell R}. 
\end{eqnarray}
The restriction from the $A$ transformation is
\begin{eqnarray}
\lambda_A&=&X_L^\dagger A_\Phi^T \lambda_A X_{\ell R} .
\end{eqnarray}

$\lambda_A$ has the same restrictions as $S^* \lambda_B$, so they can
be equal: $\lambda_A= S^* \lambda_B$.
The mass matrix $M_{\ell A}$ is a linear combination of $\lambda_{A}^{1,2,..,n}$, therefore it is also a linear combination of $\lambda_{B}^{1,2,...,n}$. 
If the symmetries do not dictate the ratios between the Higgs VEVs (e.g. for two Higgs fields $\frac{<\phi^0_1>_B}{<\phi^0_2>_B}$ and $\frac{<\phi^0_1>_A}{<\phi^0_2>_A}$ are unfixed) then $M_{\ell B}$ can be any linear combination of $\lambda_B$, and $M_{\ell B}$ can be any linear combination of $\lambda_A$. Therefore $M_{\ell B}$ has the same restrictions from the symmetry as $M_{\ell A}$, and the different equivalent representations make the same predictions.

If there is a relationship between the VEVs it is possible to get extra zeros in the mass matrices. This occurs when the $ij$th elements of the $\lambda$ matrices are non-zero, but $M^{ij}_{\ell}=<\phi^0>^T \lambda^{ij}=0$, due to a special relationship between the VEVs and elements in the $\lambda$ matrices.      
Also if one of the VEVs from a particular representation is equal to zero, then it is also possible for more zeros to be created in the mass matrix.

The neutrino mass matrices are also unchanged by a change of representation . Both $M_{\nu A}$ and $M_{\nu B}$ are linear combinations of the same $\kappa$ matrices. 
The dimension-5 Higgs-neutrino coupling has
\begin{eqnarray}
\kappa_B&=&X_L^\dagger B_\Phi^\dagger \kappa_B B_\Phi^* X_L^*\\
&=&X_L^\dagger S^\dagger A_\Phi S \kappa_B S^T A_\Phi^* S^* X_L^*\\
(S\kappa_B S^T)&=&X_L^\dagger A_\Phi (S \kappa_B S^T) A_\Phi^*  X_L^*,
\end{eqnarray}
so $S\kappa_B S^T$ can be equated to $\kappa_A$, and $M_{\nu A}$ and $M_{\nu B}$ are both linear combinations of $\kappa_B$ matrices.
 
For Dirac neutrinos
\begin{eqnarray}
\kappa_B&=&X_L^\dagger B_\Phi^\dagger \kappa_B X_{\nu R},\\
&=& X_L^\dagger S^\dagger A_\Phi^\dagger S \kappa_B X_{\nu R},\\
(S \kappa_B)&=&X_L^\dagger A_\Phi^\dagger (S \kappa_B) X_{\nu R},
\end{eqnarray}
so $S\kappa_B$ and $\kappa_A$ can be equated, and $M_{\nu A}$ and $M_{\nu B}$ are linear combinations of the same $\kappa$ matrices. Again, if the ratios between the mass matrices are not defined by the symmetry, any linear combination is a valid mass matrix, therefore both mass matrices are constrained by the symmetry in the same way.

When there is cancellations or zero VEVs, the only change to the mass matrix is some extra zero elements.

\section{$Z_2$ symmetries generate mixing matrices that are either
  ruled out, or unconstrained}\label{z2ruledout} 

Applying a $Z_2$ Higgs transformation twice leaves the Higgs fields
unchanged, so for the mixing matrix to be allowed by the no-go
theorem of \cite{nogo1}, $X_L^2=X_{\ell R}^2=X_{\nu R}^2=\pm I$.
For diagonal Higgs transformations, the components of $A_\Phi$
will also be $1$ or $-1$.
This appendix shows that mass matrices generated by a $Z_2$ transformation have equivalent restrictions to mass matrices generated by family symmetries when the Higgs field is not transforming. These restrictions are $M_\ell=X_L^\dagger M_\ell X_{\ell R}$ for charged leptons, $M_{Dirac}=X_L^\dagger M_{Dirac} X_{\nu R}$ for Dirac neutrinos, $M_\nu=X_L^\dagger M_\nu X_L^*$ for dimension-5 neutrinos, and $M_{R}=X_{\nu R}^T M_R X_{\nu R}$ for right-handed Majorana neutrinos. These situations have been ruled out by the theorem in \cite{nogo1}.

\subsection{Assuming no cancellations and no zero VEVs}

Due to the result of appendix \ref{apphiggssim} diagonal Higgs transformations can be used, and $A_{\Phi}$ can have 1 or $-1$ as diagonal elements. 
For $A_\Phi=I$ the Higgs fields do not transform -- this is equivalent to the single Higgs field case. 
For $A_\Phi=-I$, the restrictions on the charged lepton Yukawas reduces from  $\lambda=X_L^\dagger A_\Phi^T \lambda X_{\ell R}$ to $\lambda= - X_L^\dagger \lambda X_{\ell R}$ - which is equivalent to a single Higgs doublet scenario where the right-handed Higgs fields transform with $-X_{\ell R}$. The dimension-5 neutrino restrictions do not change and the Dirac neutrino Yukawa restrictions are the same as the single Higgs case when the right handed neutrinos transforms under $-X_{\nu R}$.

When $A_\Phi$ is made up of both $1$s and $-1$s, the restrictions on the charged lepton mass matrix are (from Eq.(\ref{eqchrestdiag}))
$M_\ell^{ij}=0$ unless $(X_L^{\dagger})^{ii}X_{\ell R}^{jj}=1$ or $-1$. 

This condition holds for all $i$ and $j$, so the charged lepton mass matrix is unrestricted by the symmetry. The restrictions on a dimension-5 neutrino mass matrix are similar (from Eq. (\ref{eqdim5diag}));
 $M_\nu^{ij}=0$ unless $  (X_L^{\dagger})^{ii}X_L^{* jj}=1$ or $-1$,
which also will hold for all $i$ and $j$, meaning that the neutrino mass matrix is also unrestricted by the symmetry.

Dirac neutrinos will also be unrestricted, and as a result, the seesaw neutrinos will be unrestricted by the symmetry. 

\subsection{Including the possibility of cancellations}

\subsubsection{Charged leptons}

Cancellations in the charged lepton mass matrix means that for some $i$, $j$,
$M_{\ell ij}=\lambda^1_{ij}<\Phi_1>+\lambda^2_{ij}<\Phi_2>+...=0$, while the $\lambda_{ij}$ are non-zero. This will only occur for particular $ij$'s where there is a certain relationship between the $\lambda$'s, and the VEVs. For $Z_2$ Higgs transformations there can be only two possible relationships: 
\begin{itemize}
\item $\lambda_{ij}=A_\Phi^T \lambda_{ij}$, which occurs when $(X_L^{\dagger})^{ii}X_{\ell R}^{jj}=1$. If there is a cancellation for this Yukawa relationship, the cancellations add zeros into the mass matrix in exactly the same way as the condition $M_\ell=-X_L^\dagger M_\ell X_{\ell R}$. 
 
\item $\lambda_{ij}=-A_\Phi^T \lambda_{ij}$, which occurs when $(X_L^{\dagger})^{ii}X_{\ell R}^{jj}=-1$. If this relationship led to a cancellation, the restrictions on the mass matrix would be identical to the conditions from  $M_\ell=X_L^\dagger M_\ell X_{\ell R}$.
\end{itemize}

\subsubsection{Dimension-5 neutrinos}

If there is cancellation in the neutrino mass matrix, the cancellation will occur for one of two relationships between the coupling matrices. 
\begin{itemize}
\item Cancellation when $\kappa^{ij}=A_\Phi^\dagger \kappa^{ij} A_\Phi^*$, is equivalent to the restriction $M_\nu=-X_L^\dagger M_\nu X_L^*$.
\item  Cancellation when $\kappa^{ij}=-A_\Phi^\dagger \kappa^{ij} A_\Phi^*$, is equivalent to the restriction $M_\nu=X_L^\dagger M_\nu X_L^*$.

\end{itemize}

All combinations of charged lepton mass matrix restrictions and neutrino mass matrix restrictions are the same as restrictions for single Higgs field cases.

\subsubsection{Dirac neutrinos}
 The two relationships between the Yukawa coupling terms are $\kappa^{ij}=\pm A_\Phi \kappa^{ij}$.

\begin{itemize}
\item Cancellation when $\kappa^{ij}=+ A_\Phi \kappa^{ij}$  is equivalent to $M_\nu=-X_L^\dagger M_\nu X_{\nu R}$.
\item Cancellation when $\kappa^{ij}=- A_\Phi \kappa^{ij}$ is equivalent to the restriction $M_\nu=X_L^\dagger M_\nu X_{\nu R}$.
\end{itemize}
All combinations of neutrino and charged lepton mass matrix restrictions are equivalent to single Higgs cases, in the same way as the dimension-5 neutrinos.

\subsubsection{Seesaw neutrinos}

The cancellation in the Dirac neutrino mass matrix is just the same as above, the heavy Majorana mass matrix is restricted by the symmetry by $M_M=X_{\nu_R}^\dagger M_M X_{\nu_R}^*$. The constraints from all cancellation possibilities are identical to restrictions from single Higgs doublet cases .

\subsection{Zero VEVs}

If the Higgs transformation is diagonal, and $<\phi^0_1>=0$, then the mass matrix is a linear combination of $\lambda^{2,3,...n}$, which obey $A_{\Phi}^{\alpha} X_L^\dagger \lambda^{\alpha} X_{\ell R}=\lambda^\alpha$, where $\alpha$ is the Higgs family index ranging from $2$ to $n$. These restricitons are identical to a case with one fewer higgs field. 

For a non-diagonal Higgs transformation, the different $\lambda$ matrices are related by
$\lambda=X_L^\dagger A_\Phi^T \lambda X_{\ell R}$. For $Z_2$ transformations this reduces to $\lambda_{ij}=A_\Phi^T \lambda_{ij}$ if $(X^\dagger_{L})_{ii}X_{\ell R}^{jj}=+1$, and $\lambda_{ij}=-A_\Phi^T \lambda_{ij}$ if $(X^\dagger_{L})_{ii}X_{\ell R}^{jj}=-1$ -- these are just two sets of simultaneous equations.  If the equations allow one or more of $\lambda^{2,3,..,n}$ to be non-zero, the $M_{\ell ij}$ is non-zero, otherwise the element of the mass matrix will be zero.  

There are 4 possibilities: 
\begin{itemize}
\item $M_{\ell ij}=0$ if $(X^\dagger_L)_{ii} X_{\ell R jj}=1$ otherwise  $M_{\ell ij}$ is unrestricted. This is the same restriction as $M_{\ell}=-X_L^\dagger M_{\ell} X_{\ell R}$.

\item  $M_{\ell ij}=0$ if $(X^\dagger_L)_{ii} X_{\ell R jj}=-1$ otherwise  $M_{\ell ij}$ is unrestricted. This is the same restriction as $M_{\ell}=+X_L^\dagger M_{\ell} X_{\ell R}$.

\item $M_{\ell ij}=0$ for all $(X^\dagger_L)_{ii} X_{\ell R jj}$.

\item $M_{\ell ij}$ is unrestricted for all $(X^\dagger_L)_{ii} X_{\ell R jj}$.
\end{itemize}

For Dirac neutrinos the situation is similar. The $\kappa_{Dirac}$ matrices are related by $\kappa_{Dirac}=X_L^\dagger A_\Phi^\dagger \lambda X_{\nu R}$. The same four possible restrictions arise:  $M_{Dirac}=+X_L^\dagger M_{Dirac} X_{\nu R}$, $M_{Dirac}=-X_L^\dagger M_{Dirac} X_{\nu R}$, $M_{Dirac}$ is unrestricted, and $M_{Dirac}=0$.

For neutrinos with masses due to a dimension-5 operator, the $\kappa$ matrices are related by $\kappa=X_L^\dagger A_\phi^\dagger \kappa A_\phi^* X_L^*$. The possible restrictions on the neutrino mass matrix are:  $M_{\nu}=+X_L^\dagger M_{\nu} X_L^*$, $M_{\nu}=+X_L^\dagger M_{\nu} X_L^*$, $M_{\nu}$ is unrestricted, and $M_{\nu}=0$.

A right-handed Majorana mass matrix is unaffected by the VEVs, so the usual restriction applies: $M_R=X_{\nu R}^T M_R X_{\nu R}$. 

The combinations of neutrino and charged lepton mass matrix restrictions give four possibilities: The restrictions are identical to single Higgs field cases, the neutrinos are massless, the
charged leptons are massless, or the mixing matrix is unconstrained by the symmetry. Therefore a $Z_2$ transformation predicts mixing matrices that are either ruled out or unconstrained by the symmetry. 

Note that this does not mean that groups which have $Z_2$ as a subgroup can be
ruled out. If the $Z_2$ subgroup gives mixing matrices that are not
allowed, then the group is ruled out. However, if the $Z_2$ subgroup
leaves the masses and mixing angles
unrestricted, then the group is still allowed.

\bibliography{nhiggs}

\end{document}